\documentclass[%
 reprint,
superscriptaddress,
 amsmath,amssymb,
 aps,
prb,
]{revtex4-1}

\usepackage{graphicx}% Include figure files
\usepackage{dcolumn}% Align table columns on decimal point
\usepackage{bm}% bold math

\begin{document}

% \preprint{APS/123-QED}
% l
\title{Quasiparticle Screening near a Bosonic Superconductor-Insulator Transition \\Revealed by Magnetic Impurity Doping}
% Force line breaks with \\

% .\\Evidence of Screening by Virtual Quasiparticles Near a Bosonic Superconductor to Insulator Transition from Magnetic Impurity Doping Studies\\Evidence of Virtual Quasiparticle Screened Coulomb Interactions in the Cooper Pair Insulator\\Magnetic Impurity Doping of a Cooper Pair Insulator Reveals Evidence of Virtual Quasiparticle Screened Coulomb Interactions\\, Cooper Pair Insulator Transport Appears Enhanced by Quasiparticle Screening
\author{Xue Zhang}
\author{James C. Joy}
\author{Chunshu Wu}
\affiliation{Department of Physics, Brown University, Providence, RI, USA }

\author{Jin Ho Kim}
\author{{J. M. Xu}}
\affiliation{School of Engineering, Brown University, Providence, RI, USA }

\author{James M. Valles, Jr.}
\affiliation{Department of Physics, Brown University, Providence, RI, USA }

\date{\today}% It is always \today, today,

\begin{abstract}
Experiments show that the Cooper pair transport in the  insulator phase that forms at thin film superconductor to insulator transitions (SIT) is simply activated.
This activated behavior depends on the microscopic factors that drive the localization of the Cooper pairs. To test proposed models, we investigated how a perturbation that weakens Cooper pair binding, magnetic impurity doping, affects the characteristic activation energy, $T_0$.   The data show that $T_0$ decreases monotonically with doping in films tuned farther from the SIT and increases and peaks in films that are closer to the SIT critical point. These observations provide strong evidence that the bosonic SIT in thin films is a Mott transition driven by Coulomb interactions that are screened by virtual quasi-particle excitations.  This dependence on underlying fermionic degrees of freedom distinguishes these SITs from those in micro-fabricated Josephson Junction Arrays, cold atom systems, and likely in high temperature superconductors with nodes in their quasiparticle density of states.  
% This approach was partially motivated by the fruitfulness of early investigations of the different responses of superconductors to non-magnetic and magnetic impurities.  Those led to microscopic models of the detrimental effects of spin flip scattering on the Cooper pairing energy. 
\end{abstract}

\maketitle
The activation energy characterizing a process in a condensed matter system provides a window into its quantum many body ground state.  For example, the resistance of fractional quantum hall states decreases with decreasing temperature at a rate dictated by the energy to create spatially separated quasiparticle-quasihole pairs out of the Laughlin ground state\cite{Laughlin1983,Boebinger1987,Girvin1985}.  Similarly, the heat capacity of BCS superconductors decreases at a rate dictated by the binding energy of electrons in Cooper pairs\cite{Tinkham1996a}.  This paper focuses on an activation energy, $T_0$,\cite{Sambandamurthy2004a,Baturina2008,Lin2014a} that grows from zero at disorder tuned bosonic superconductor to insulator quantum phase transitions in ultra-thin films \cite{Stewart2007}.  It is a barrier to Cooper pair tunneling between localized states\cite{Stewart2007}.  There are competing models for the physical origin of this barrier.  Some attribute it to disorder induced Anderson localization effects\cite{Bouadim2011,Gangopadhyay2013,Loh2016} and others to repulsive Coulomb interaction effects\cite{Dubi2006,Beloborodov2007,Fistul2008}.  Results from experiments to date, which have shown that $T_0$ depends on magnetic field\cite{Sambandamurthy2004a,Baturina2007a,Nguyen2009a,Lin2014a}, magnetic frustration\cite{Nguyen2009a} and normal state resistance\cite{Steiner2005a,Stewart2007}, can be accounted for by many of the models leaving the microscopic origins of $T_0$ mysterious.  Here, we present measurements of the dependence of $T_0$ on magnetic impurity doping, which weakens Cooper pairing and magnetic frustration, which alters Cooper pair tunneling rates. We describe how the results indicate that the activation barrier depends directly on the average Cooper pair binding energy.  Such a dependence arises in Cooper pair tunneling transport models that include screening of a Coulomb barrier by virtual quasi-particle excitations\cite{Beloborodov2007}.  

Models for Cooper pair localization at the SIT have led to predictions for their activated resistance, 
\begin{equation}
R=R_0\exp(T_0/T)
\label{RT}
\end{equation}
In most, $T_0$ results from a competition between either potential disorder or Coulomb interactions that localize pairs and pair tunneling, characterized by a hopping rate $t$ or a Josephson coupling energy, $E_J$, that delocalizes pairs.  Potential disorder drives Anderson localization\cite{Bouadim2011,Gangopadhyay2013} of pair states with energies below a mobility edge in the density of states. The activation energy corresponds to the gap between localized and mobile pair states\cite{Gangopadhyay2013}.  It increases with disorder or decreasing $t$. Coulomb interactions, on the other hand, drive a Mott transition by creating a blockade to pair motion between localized states\cite{Efetov}.  The blockade is characterized by a charging energy, $E_c=2e^2/C$, that depends on the capacitance between a localized state and its environment\cite{Efetov,Cha1994, Beloborodov2007, Fistul2008, Syzranov2009, Feigelman2010,Baturina2013,Swanson2014}.  In the limit, $E_c\gg E_J$, a Mott gap appears in the transport, $T_0\approx E_c(1-\frac{\beta E_J}{E_c})$, in which the second term arises from screening by Cooper pair motion\cite{Fazio1991,Beloborodov2007} with $\beta$ a coordination number dependent constant.  Experiments that measure how $T_0$ responds to changes in parameters like $E_J$ can test these models and thus, yield insight into Cooper pair localization and transport. 

\begin{figure}[h!]
\centering
\includegraphics[width=1.0\linewidth]{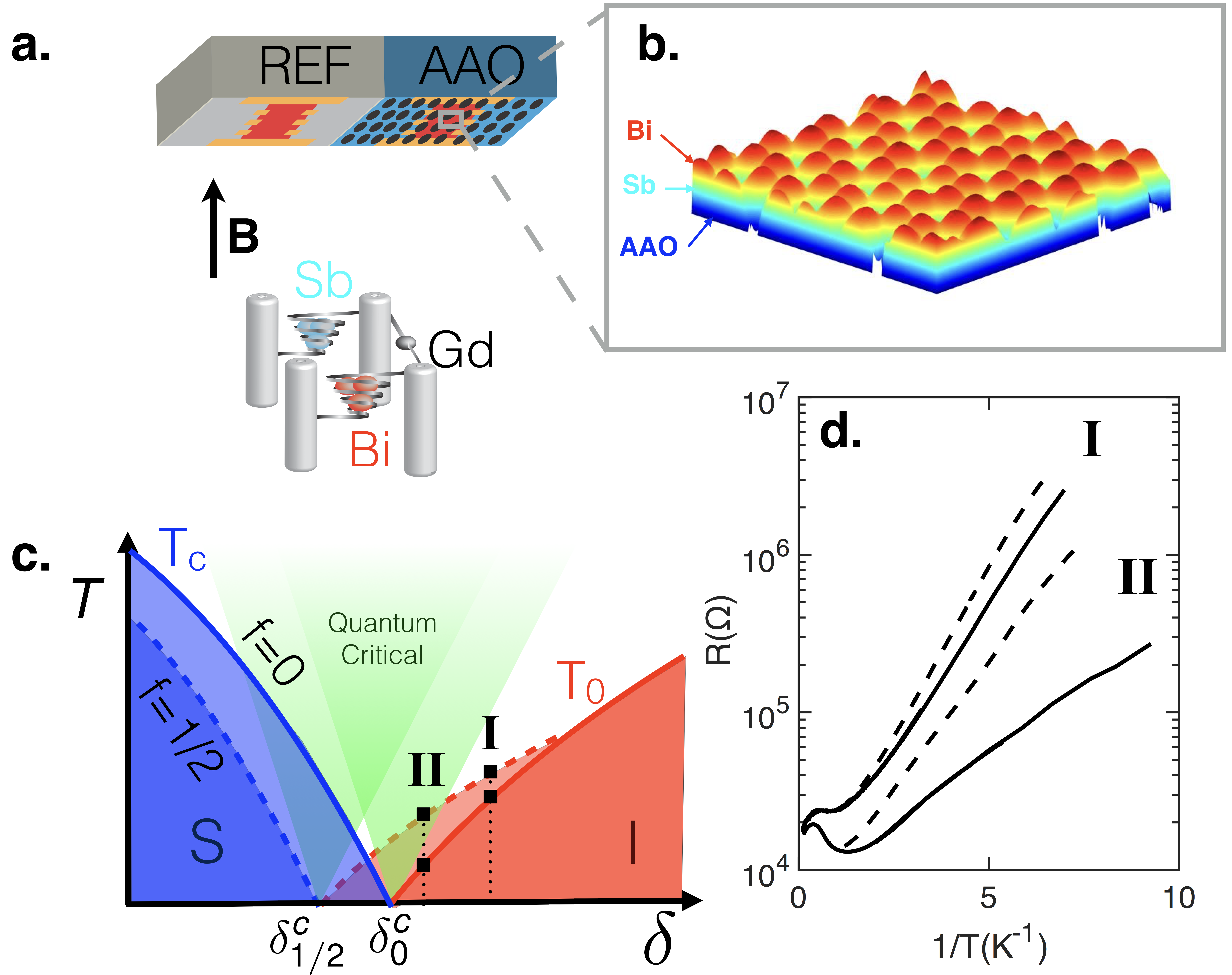}
\caption{a) Sketch of experimental set-up displaying  side by side flat glass (REF) and AAO substrates positioned over the Sb, Bi, and Gd evaporation sources in a magnetic field {\bf B} directed as shown.  b) Atomic force microscopy (AFM) image of AAO falsely colored to indicate the evaporation layers, substrate height variations and 100 nm period nanopore array. c) Schematic phase diagram of temperature vs. coupling constant, $\delta\sim R_N$ for a superconductor to Cooper Pair Insulator quantum phase transition. I and II refer to the films investigated. There is a critical point for each of the frustrations, $f=0$ and $1/2$. d) Sheet resistance on a logarithmic scale versus inverse temperature for undoped films, I and II, at $f=0$ (solid lines) and $f=1/2$ (dashed lines). 
}
\label{fig:afm}
\end{figure}

We have employed a thin film platform\cite{Stewart2007} that enables unique methods for probing the origins of $T_0$. The films can be systematically doped with magnetic impurities, which reduces the Cooper pair binding energy, $2\Delta$ and can be subjected to magnetic frustration, which reduces the average Josephson coupling between localized regions\cite{Nguyen2009a} (see Figs. \ref{fig:afm} a,b).  Since $E_J\propto \Delta$, the doping also reduces $E_J$.  For both the Anderson and Mott models, reducing $E_J$ is expected to enhance $T_0$ and thus, Cooper pair localization.  Surprisingly, we found that while magnetic frustration always enhances $T_0$, magnetic impurity doping can reduce $T_0$.  We discuss how this result intimates that the superconductor to Cooper pair insulator transition is a Mott transition with a Coulomb blockade energy that depends on the pair binding energy.  

Sub-nanometer thick amorphous Bi films were fabricated and measured {\it in situ} in the UHV environment of a dilution refrigerator based evaporator.  Bi vapor was quench condensed onto an Sb wetting layer on the surface of two substrates simulataneously: an Anodized Aluminum Oxide substrate, which has regular height variations and an array of pores, and a flat, fire polished glass substrate.  Both substrates were held at a temperature, $T=T_s\approx 10$ K within the UHV environment of a dilution refrigerator cryostat (Fig. \ref{fig:afm}). The depositions of Bi and Sb were measured using a quartz crystal micro-balance. The Cooper pair insulator state forms in films on AAO substrates because of nanometer scale height variations, $h(x)$ on the AAO surface (see Fig. 1b)\cite{Hollen2011}.  These lead to local surface slope variations that produce film thickness variations $d(x)$:

\begin{equation}
d(x)=\frac{d^{\rm dep}}{\sqrt{1+(\nabla h(x))^2}}
\end{equation}

Since $T_c$ depends on film thickness\cite{Valles1989} the thickness variations correspond to coupling constant inhomogeneities that localize Cooper pairs to form the Cooper pair insulator state. Insulating films on flat substrates, by contrast,  consist only of localized, unpaired electrons\cite{Finkelstein1994} \cite{VallesJr.DynesR.C.andGarnoJ.P.1992, hollen2013b}. The film on the flat substrate served as a reference for monitoring 1) the maximum thickness and pairing amplitude that could appear in the films deposited on AAO and 2) the pair breaking effects of the magnetic impurity depositions.

Film sheet resistances were measured as a function of temperature, $R(T)$ \textit{in situ} using standard four-point ac and dc techniques with sufficiently low current bias($0.2$ nA) to ensure that the measurements were performed in the linear portion of the  current-voltage characteristics. A superconducting solenoid applied magnetic fields perpendicular to the films. 

The array of pores in the films produced by the AAO templates enable us to explore magnetic field induced frustration effects on the Cooper pair insulator phase.  In fact, the appearance of oscillations in the magnetoresistance was an early direct sign of localized Cooper pairs in a thin film system \cite{Stewart2007}. The activation energy and location of the SIT critical point (see Fig. \ref{fig:afm} c) is periodic in the frustration $f=H/H_M$, where $H_M$ is the magnetic field that produces one superconducting flux quantum per plaquette.  $H_M=0.21 T$ for the 100 nm average center to center spacing of nearest neighbor pores. This frustration dependence can be attributed to a modulation of the average Josephson coupling between islands with a period of one flux quantum per plaquette\cite{Nguyen2009a,Muller2013a}. The average appears as $<E_J>=E_{J0} <\sum_{<i,j>}\cos(\phi_i-\phi_j-A_{ij})>$ in Quantum Rotor Models, where $\phi_i$ and $\phi_j$ are the phases on neighboring islands and $A_{ij}$ is the line integral of the vector potential between islands. For a honeycomb array of islands, the energy barrier for Cooper pair transport is highest for $f=1/2$\cite{Granato2016a}. Phenomenologically, $<E_J(f)>\propto E_{J0} F(2\pi f)$, where $F$ is a periodic function.

Magnetic impurity doping involved depositing Gd atop the Cooper pair insulator film\cite{Parker2006}. The impurities produce time reversal symmetry breaking spin flip scattering, which reduces the pair binding energy $2\Delta$. Their effect extends uniformly through the entire thickness of the films since the films are much thinner ($d\le 1$ nm) than the superconducting coherence length ($\xi \ge 10$ nm)\cite{Chervenak1995}. The Gd deposition amounts, $x_{Gd}$ which were below the micro-balance resolution, were monitored using a calibrated timing method and by measuring their effects on the $T_c$ of the reference film.  The two methods agreed well.  In the following, we use the relative $T_c$ shift
\begin{equation}
\alpha_{Gd}=1-T_c(x_{Gd})/T_c(0)
\end{equation}
to represent doping amount.  The estimated maximum Gd doping in these experiments corresponded to $< 0.03$ monolayers.

We studied the effects of magnetic impurity doping and magnetic frustration on two films, I and II, that had different activation energies to explore how proximity to the SIT critical point influences the response.  Points for films I and II are indicated on the schematic phase diagram in Fig. 1c. according to their relative activation energies obtained from fits to the data shown in Fig. 1d. Other film I and II parameters are in the Table.  The phase diagram shows two distinct critical points for the two frustrations investigated, $f=0$ and $f=1/2$\cite{Granato2018}.  The tuning parameter $\delta$, corresponds to either $1/d$ or $R_N$, where $R_N$ is sheet resistance measured at 8K.  Previous work\cite{Stewart2007} indicated that the critical values of the tuning parameters for the SIT followed $\delta_{0}^c>\delta_{1/2}^c$.   
% xz Tc_ref(I) = 2.59K. Tc_ref(II)=2.92K is another indicator that II is closer to the transition.
\begin{table}
\caption{\label{tab:table2} Film I and II parameters.}
\begin{ruledtabular}
\begin{tabular}{lccccc}
 &$R_N$&$d_{Bi}$&$T_0(0)$  &$T_0(1/2)$  &$T_c(0)$\\
\hline
I & 18.6 $k\Omega$ & 0.99 nm& 0.86 K & 0.98 K & 2.59 K$$ \\
II & 16.7 $k\Omega$ & 1.2 nm & 0.40 K & 0.75 K & 2.92 K$$\\
\end{tabular}
\end{ruledtabular}
\end{table}
\begin{figure}[h!]
\centering
\includegraphics[width=1.0\linewidth]{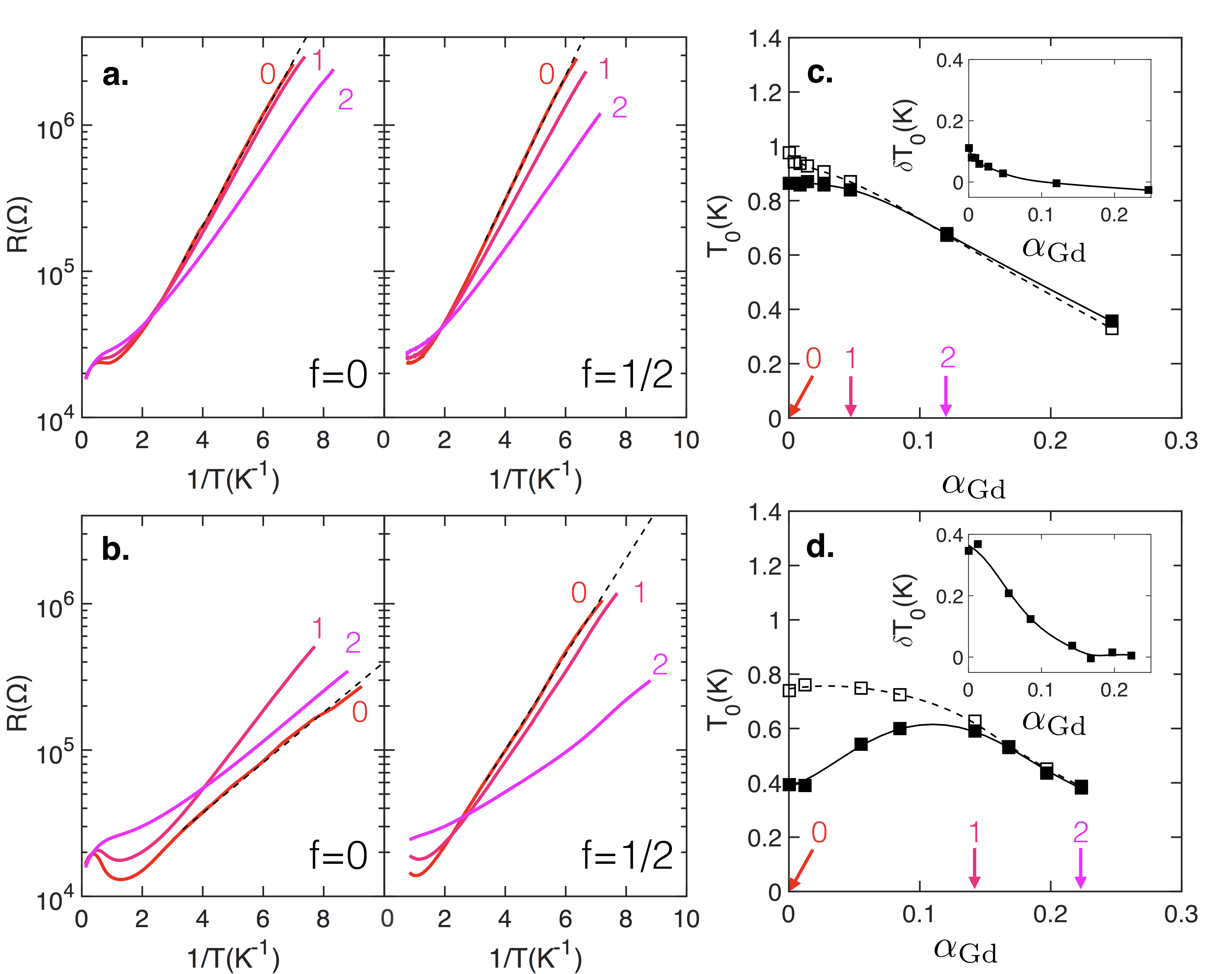}
\caption{Magnetic Impurity Doping Response of Resistance and Activation Energy. $\log R$ versus $1/T$ for a) film I and b) film II at $f=0$ and $f=1/2$, respectively. Three representative curves are displayed in each panel. The numerical labels increase with Gd doping, where 0 represents no doping. The dashed lines give examples of fits to Eq. \ref{RT} in the low temperature regions. c),d) Activation energies obtained from linear fits to $R(T)$ data as in a) as a function of magnetic impurity induced pairbreaking for films I and II, respectively.  The pairbreaking was obtained from the relative $T_c$ shift on the reference film as described in the text. The filled squares correspond to $f=0$ and open squares are $f=1/2$. The lines are guides to the eye.  Insert: $\delta T_0=T_0^{f=1/2}-T_0^{f=0}$ as a function of doping induced pairbreaking.
}
\label{fig:RT_T0}
\end{figure}
The effects of Gd doping on I and II are displayed in Fig. \ref{fig:RT_T0}.  The Arrhenius plots  show that the activated form of $R(T)$ persists through the doping range.  The evolution of $T_0$ with low doping depends on distance from the SIT.  $T_0$ decreases or remains nearly constant and then decreases with Gd doping in the three films farthest from the SIT.  By contrast, the film closest to the SIT exhibits a maximum in $T_0$.  At the higher doping levels, $T_0$ becomes independent of frustration and decreases with doping.  The  difference between the activation energies at the two frustrations, $\delta T_0=T_0^{1/2}-T_0^0$, is larger for the film closer to the SIT.  For both films, $\delta T_0$ goes to zero, nearly linearly, at higher doping levels.  

We break the discussion of the primary experimental results shown in Figs. \ref{fig:RT_T0}b and d into two parts focusing first on the low doping regime where $T_0$ depends on frustration.  This frustration dependence indicates that Cooper pairs are the dominant charge carriers \cite{Hollen2014} in this regime.  In both films I and II, $T_0(1/2)>T_0(0)$, which is consistent with all of the activation energy models discussed above.  Introducing frustration reduces the average $E_J$ or $t$ or Cooper pair tunneling rate.  This reduction increases the mobility gap\cite{Gangopadhyay2013} or reduces the Cooper pair screening of the Coulomb blockade energy to increase $T_0$\cite{Fazio1991,Beloborodov2007}.  

On the other hand, $T_0$'s decrease with doping for three cases (film I at f=0 and f=1/2, and film II at f=1/2) opposes simple expectations.  Pairbreaking reduces $2\Delta$, which should again reduce $E_J$ or $t$ to make $T_0$ rise as with frustration. Similarly if the impurities were to randomly transform links into $\pi$ junctions\cite{Bulaevskii1977}, their effect would be to reduce $E_J$ or $t$ to make $T_0$ rise at large doping\cite{Granato2017}.  Thus, these three cases rule out disorder induced localization models in which $t$ is the only $\Delta$ dependent parameter\cite{Gangopadhyay2013}.  They also rule out Coulomb interaction models in which the charging energy depends only on the geometry of the localized states and the dielectric constant of the intervening insulating material\cite{Fistul2008}.  Magnetic impurity doping is not expected to influence the electric polarizability of the intervening material.  It might influence the geometry by causing the islands to shrink.  That effect, however, would increase charging energies and thus, $T_0$.
%xz Magnetic impurity does not influence dielectric properties. Is it simply because the concentration is low and electrical field is not affected in normal state?

A possible explanation for $T_0$ decreasing with pairbreaking is that the charging energy depends directly on $\Delta$.  Such a dependence emerges in small single Josephson junctions when the bare charging energy, $E_{C0}$, which is set by the island geometry and tunnel barrier dielectric constant, greatly exceeds $\Delta$, i.e. $E_{C0} \gg\Delta$\cite{Larkin,Ambegaokar1982,Chakravarty1987,Beloborodov2007}.  In that limit,  virtual inter island tunneling of quasiparticles above the energy gap $\Delta$ renormalizes the capacitance of single junctions\cite{Ambegaokar1982} or effectively screens the Coulomb interaction\cite{Chakravarty1987}.  Beloborodov and coworkers\cite{Beloborodov2007} included this effect in a model of granular films that had been considered previously\cite{Chakravarty1987} to derive a renormalized charging energy:
\begin{equation}
E_C=\frac{2\Delta}{3\pi^2 g} \ln\left(gE_{C0}/\Delta\right)
\label{EC}
\end{equation}
where $g=G/(2e^2/\hbar)$ is the dimensionless normal state conductance between grains. With this form for the charging energy, $T_0$ becomes:
\begin{equation} \label{T0B}
T_0=\frac{2\Delta}{3\pi^2 g} \ln\left(gE_{C0}/\Delta\right)-\beta g\Delta F(2\pi f)
\end{equation}
for films close enough to the SIT that Cooper pair screening reduces $T_0$ through the Josephson energy term $E_J \propto g\Delta$.  A rough estimate of $E_{C0}$ for the dots that form in films on AAO suggests that this model applies here.  The dots have characteristic diameters below $30 $ nm, which gives $E_{c0} \approx 200 $K for a dielectric constant of 10, which greatly exceeds $\Delta$.  

\begin{figure}[h!]
\centering
\includegraphics[width=0.9\linewidth]{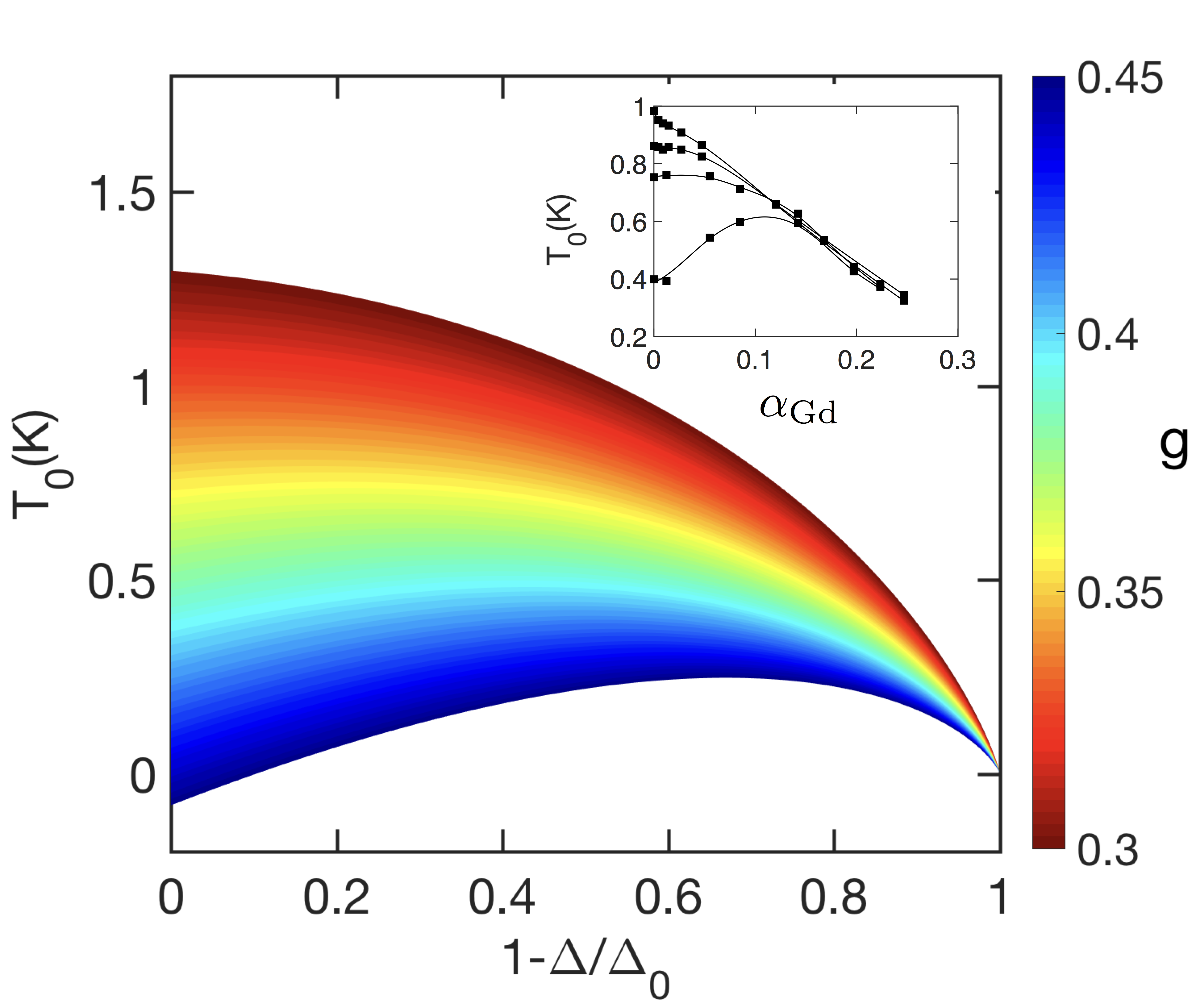}
\caption{Comparison of the data with the Mott Insulator with Virtual Quasiparticle Screening Model.  The main figure gives contours of $T_0$ at constant $g$ versus the normalized change in $\Delta$ as calculated from Eq.(\ref{T0B}) for the case $f=0$.  The normalized change in $\Delta$ is proportional to $\alpha_{Gd}$ for low doping. The model parameters were set to $E_{C0}=200$ K, $\beta=1$, and $\Delta_0=5$ K with $g$ varying from 0.3 to 0.45. Over this range of $g$ the activation energy at zero doping changes sign indicating that it spans the SIT. Using $f=1/2$ generates a qualitatively similar spectrum of curves that have a weaker maximum at the largest $g$. Inset: Experimental results for films I and II with lines that are guides to the eye.
}
\label{fig:model}
\end{figure}

Model predictions embodied in Eq. \ref{T0B} qualitatively capture the experimental results in the low doping regime.  Fig. \ref{fig:model} displays contours of $T_0$ at fixed $g$ or distance from the SIT critical point as a function of $\delta\Delta=1-\Delta/\Delta_0$ where $\Delta_0$ is the zero doping energy gap.   The contours were calculated using parameter values, $E_{C0}=200$ K, $\beta=1$, $\Delta_0=5$ K and $f=0$ with $g$ varying from 0.3 to 0.45, that fell in a plausible range for these AAO films.  At lower $g$, farther from the SIT $T_0$ decreases monotonically with $\delta\Delta$.  At higher $g$, closer to the SIT, $T_0(\delta\Delta)$ develops a maximum.  These trends qualitatively resemble the observed trends in the $T_0$ data as a function of $\alpha_{Gd}$ and distance from the SIT critical point (see inset of Fig. \ref{fig:model}).This agreement goes a long way toward establishing this superconductor to insulator transition as a Mott transition. The screening effect differentiates it from the Mott transition that occurs in micro-fabricated Josephson Junction Arrays for which there is little quasiparticle screening because $\Delta > E_c$\cite{Delsing1994}.  Similarly, it differs from cold atom system Mott transitions because those bosons cannot decompose into constituent parts\cite{Greiner2002}.

A vexing question has been why does Cooper pair insulator transport appear simply activated in disordered films?  The disorder suggests that variable range hopping models, which produce a fractional power (e.g. 1/2) in the activation energy exponent\cite{Beloborodov2007}, should apply. The fractional power appears when electrons must hop between localized states that are randomly distributed in energy and space.  The granular model for transport with quasi-particle screening embodied by the activation energy in Eq. (5) suggests that the distribution of energy levels is more uniform in the CPI phase.  If the grains or dots that serve as the localization sites are large enough that $\Delta$ assumes a nearly constant value across a sample, then the logarithmic dependence on $E_c$ makes the activation energy insensitive to variations in the size of the grains.  Thus, the distribution of activation energies can be smaller than one might expect for these disordered systems.

It is also interesting to consider the implications of the present results for the bosonic SITs in high temperature superconducting cuprates\cite{Bollinger2011}.  The nodes in their superconducting density of states will affect the properties of a Cooper pair insulator phase.  The availability of low energy states could make the virtual quasiparticle screening even more effective than in s-wave systems, which would tend to reduce their activation energies.  It may also alter the power in the exponent that characterizes the activated transport in the same way the penetration depth temperature dependence depends on the structure of the gap\cite{Hardy1993}.

Finally, the disappearance of $T_0$'s frustration dependence at higher doping levels signals a crossover from transport that involves Cooper pairs to quasiparticle dominated transport.  The crossover is smooth: the  $R(T)$ (Figs. 2a,c) maintain an activated form and $T_0$ evolves without any clear discontinuities in its value or slope.  The continued decrease of $T_0$ with Gd doping suggests that the quasi-particle transport depends directly on $\Delta$. This dependence can arise if the transport is quasi-particle tunneling between superconducting dots as proposed to explain negative magneto-resistance in granular Pb \cite{Barber2006} and Indium Oxide films \cite{Gantmakher1998}. Within this model, the inferred values of $2\Delta$ at the crossover, presuming $T_0=2\Delta$, are 0.83 K and 0.6 K for films I and II, respectively. Both of these values fall below the transition temperatures of their associated reference films, which makes them reasonable.   

To summarize, we investigated the influence of magnetic impurity doping, magnetic frustration, and sheet resistance on the transport of the Cooper pair insulator phase in amorphous Bi films on AAO substrates.   The response of the transport activation energy, $T_0$, to changes in these variables is consistent with it being proportional to $\Delta$, the energy binding the Cooper pairs. This dependence agrees with a model\cite{Beloborodov2007} in which virtual quasi-particle tunneling processes screen the Coulomb interactions that impede boson tunneling transport while ruling out a number of others\cite{Dubi2006,Fistul2008,Bouadim2011,Gangopadhyay2013,Loh2016}.  The observations distinguish the Cooper pair insulator phase that develops in disordered s-wave films from that in micro-fabricated Josephson Junction arrays\cite{Delsing1994} and the bose insulator phase in cold atom systems\cite{Greiner2002} in which virtual quasiparticle processes exert negligible influence.   

\section*{Acknowledgements}
We are grateful to S. Kivelson, N. Trivedi, M. Mueller, and especially, I. Beloborodov and E. Granato for helpful discussions.  This work was partially supported by NSF Grant Nos. DMR-1307290 and DMR-1408743 and the AFOSR.

\bibliography{SIT_6-21-2018.bib}
\end{document}